\documentclass[12pt]{iopart}
\usepackage{graphicx,nowidow,times,epstopdf,color}

\usepackage{iopams}  

\expandafter\let\csname equation*\endcsname\relax

\expandafter\let\csname endequation*\endcsname\relax

\usepackage{amsmath,bm}

\newcommand{\ket}[1]{\left|{#1}\right>}
\newcommand{\bra}[1]{\left<{#1}\right|}

\newcommand{\opinner}[3]{\left<{#1}\vphantom{#1}\vphantom{#3}\right|{#2}\left|{#3}\vphantom{#1}\vphantom{#3}\right>}
\newcommand{\rvec}[1]{\pmb{#1}}
\newcommand{\dyadic}[1]{\pmb{#1}}
\newcommand{\bcdot}{\bm{\cdot}}
\newcommand{\D}{\mathrm{d}}
\newcommand{\I}{\mathrm{i}}
\newcommand{\TP}[1]{{#1}^\mathrm{\,\textsc{t}}}
\newcommand{\E}[1]{\mathrm{e}^{\mbox{\footnotesize$#1$}}}

\newcommand{\DET}[1]{\det\!\left\{#1\right\}}

\newcommand{\ML}[1]{\widehat{#1}_\textsc{ml}}
\newcommand{\MLo}[1]{\widehat{#1}_\textsc{ml,0}}
\newcommand{\norm}[1]{\left|\!\left|#1\right|\!\right|}
\newcommand{\FML}{\dyadic{F}_\textsc{ml}}
\newcommand{\gML}{\bm{g}_\textsc{ml}}

\newcommand{\BETAr}[2]{\mathrm{I}_{#1}\left(#2\right)}

\begin{document}

\title{Bayesian error regions in quantum estimation I: analytical reasonings}

\author{Yong Siah Teo$^{1,2}$, Changhun Oh$^1$, and Hyunseok Jeong$^1$}

\address{$^1$ Center for Macroscopic Quantum Control, Seoul National University, 08826 Seoul, South Korea}
\address{$^2$ BK21 Frontier Physics Research Division, Seoul National University, 08826 Seoul, South Korea}
\ead{ys\_teo@snu.ac.kr and v55ohv@snu.ac.kr}
\vspace{10pt}

\begin{abstract}
Results concerning the construction of quantum Bayesian error regions as a means to certify the quality of parameter point estimators have been reported in recent years. This task remains numerically formidable in practice for large dimensions and so far, no analytical expressions of the region size and credibility (probability of any given true parameter residing in the region) are known, which form the two principal region properties to be reported alongside a point estimator obtained from collected data. We first establish analytical formulas for the size and credibility that are valid for a uniform prior distribution over parameters, sufficiently large data samples and general constrained convex parameter-estimation settings. These formulas provide a means to an efficient asymptotic error certification for parameters of arbitrary dimensions. Next, we demonstrate the accuracies of these analytical formulas as compared to numerically computed region quantities with simulated examples in qubit and qutrit quantum-state tomography where computations of the latter are feasible.
\end{abstract}

%
\vspace{2pc}

\section{Introduction}

Quantum estimation or tomography with informationally complete data involves the reconstruction of a point estimator $\widehat{\rvec{r}}$ for an unknown parameter $\rvec{r}$ (generally a multivariate vectorial quantity), which may represent a quantum state, phase, expectation values of arbitrary observables, and so forth. A complete assessment of $\widehat{\rvec{r}}$ in order to perform subsequent predictions with it requires the knowledge of its corresponding measurement errors. Methods for correctly and systematically constructing error bars for scalar parameters, or \emph{error-regions} for multivariate parameters, are thus of imminent importance in scientific inquiry. 

There exist a heuristic class of methods that offer an extrapolated error analysis by taking the variance of simulated data generated from the observed dataset. This idea of ``bootstrapping'' or ``resampling''~\cite{Efron:1993bs,Davison:1997ri}, while apparently capable of economically generating error certifications for estimators, can be shown to produce nonconservative conclusions~\cite{Suess:2017np} that would misrepresent the actual statistics of the estimator. It cannot be overemphasized that proper statistical methods are required to construct meaningful error regions. As an important study, we shall analyze regions for the point estimator $\widehat{\rvec{r}}=\ML{\rvec{r}}$ that is derived from the \emph{maximum-likelihood}~(ML) strategy. Statistically, the ML estimator $\ML{\rvec{r}}$ is a parameter that is more likely to be the true one than others for a given observed dataset. Such an estimator is known to be efficiently computable with the help of proper gradient methods~\cite{Rehacek:2007ml,Fiurasek:2001dn,Shang:2017sf}.

A statistically meaningful construction of error regions for data that are \emph{actually observed}, as it turns out, is rather closely related to the theory of Bayesian inference that interprets observed data as an avenue for updating an observer's prior information about the unknown true parameter $\rvec{r}$. In recent years, Refs.~\cite{Shang:2013cc,Li:2016da} have successfully constructed optimal Bayesian credible regions $\mathcal{R}$ (or simply Bayesian regions) for ML estimators of quantum states, the so-called ML regions as coined in the references. These region possesses the smallest \emph{size} for a given \emph{credibility} with respect to observed data. In terms of their interpretations, the size quantifies how large the prior content is in $\mathcal{R}$ for which there is a certain probability (credibility) that $\rvec{r}$ lies in $\mathcal{R}$\footnote{This is a consequence of the Bayesian probabilistic viewpoint of $\rvec{r}$.}.

These Bayesian regions should be formally distinguished from the confidence regions constructed in \cite{Christandl:2012qs,Blume-Kohout:2012eb}, or their simplified variants proposed in \cite{Faist:2016eb}. The latter quantify errors with respect to \emph{all} conceivable data including those that are unobserved. No conclusion can be drawn from a single experimental run. Typically, some form of distribution over all datasets has to be expected, and in the case of cryptography for instance, this expectation becomes invalid due to the presence of eavesdropping. The former, on the other hand, derives statistical statements solely from measured data and is hence logically reliable in any setting.

For large dimensions, it has been shown that the complex structures of a convex parameter space and its boundaries render the construction of Bayesian regions generally an NP-hard problem, as is also the case for confidence regions~\cite{Suess:2017np}. In quantum-state tomography, sophisticated Monte Carlo methods have been developed and applied to sample the state space of bipartite systems with modest dimensions in order to compute the region size and credibility~\cite{Shang:2015mc,Seah:2015mc}. The certification of estimators for larger dimensions, nevertheless, remains a work in progress and thus far, no known analytical expressions are found for the size and credibility as a result of their asymptotically intractable computational complexities with the parameter dimension.

The main results of our contributions can be divided into two parts. The first part of our work supplies easy-to-calculate approximations for the size and credibility of Bayesian regions with uniform priors in the limit of large data-sample size. The expressions describe not only the case where the ML estimator $\ML{\rvec{r}}$ is an interior point in the entire parameter space, but also the case where $\ML{\rvec{r}}$ lies on its boundary. The latter case is common whenever $\rvec{r}$ is a boundary point, especially for large dimensions. These results offer an asymptotic and approximate estimate for the actual size and credibility which are useful for certifying estimators of large dimensions and sufficiently large sample size. We show, with examples of quantum-state tomography, that the expressions work well even for moderately large sample size. In the companion article~\cite{BR:part2} we shall discuss various adaptive methods that optimize tomographic accuracy in the context of these Bayesian regions.

The article is organized in the following manner. After a brief overview of the general theories of and notational introduction to quantum estimation and Bayesian regions in Sec.~\ref{sec:theories}, we shall present asymptotic analytical approximations for the size and credibility of these regions and examine their characteristics in Sec.~\ref{sec:results}. The formulas shall be derived for uniform parameter priors, and are applicable to convex parameter spaces of arbitrary dimension. Thereafter, we look at specific examples in quantum-state tomography and validate these results for the quantum state space in Sec.~\ref{sec:examples}.

\section{Basic theories and notations}
\label{sec:theories}
\subsection{Quantum estimation}
A quantum system is defined by a (generally vectorial) parameter $\rvec{r}=(r_1\,\,r_2\,\,\ldots\,\,r_d)^{\textsc{t}}$. For instance, in quantum tomography, $\rvec{r}$ would represent some quantum state of particles; in quantum metrology, $\rvec{r}=\phi$ could describe the phase of a Mach-Zehnder interferometer, and the list goes on. To characterize $\rvec{r}$, the observer measures a POM (probability operator measurement\footnote{Or more mathematically a positive operator-valued measure.}) $\sum_k\Pi_k=1$ to obtain data $\mathbb{D}$ according to the measurement probabilities $p_k=p_k(\rvec{r})$.

Based on $\mathbb{D}$, we may infer $\rvec{r}$ using standard tools in statistical inference. In particular, we focus on an important type of estimator that is ubiquitous in the discussion of core statistical topics, namely the estimator that maximizes the likelihood function $L(\mathbb{D}|\rvec{r})$---the conditional probability of gathering the data $\mathbb{D}$ given the parameter $\rvec{r}$---\emph{over some constrained parameter space} of interest (like the physical quantum state space in quantum-state tomography). In typical situations, the ML estimator $\ML{\rvec{r}}$ is unique, apart from interferometric situations~\cite{Pezze:2007mz}, for instance, where $L(\mathbb{D}|\rvec{r})$ has local minima (within the $2\pi$ period). Then, the latter case will eventually converge to the former as more independent data are collected.

In our present context, we shall consider an experimental situation where the data $\mathbb{D}$ is collected by measuring a given number of sample size or number of data copies $N$, where each copy is independent and identically distributed (i.i.d.) according to a fixed but unknown distribution given by $p_k$. The statistics of measured frequencies $\mathbb{D}=\{n_k\}$ $\left(\sum_kn_k=N\right)$ for every $p_k$ in this situation is multinomial.

\subsection{Bayesian regions}

We shall investigate two different kinds of Bayesian regions with good physical meanings. Almost no derivations of the properties for these regions are repeated in this section. Rather, important remarks about these properties are listed to set the stage for upcoming discussions.
 
\subsubsection{Credible regions}
\begin{figure}[t]
	\center
	\includegraphics[width=0.4\columnwidth]{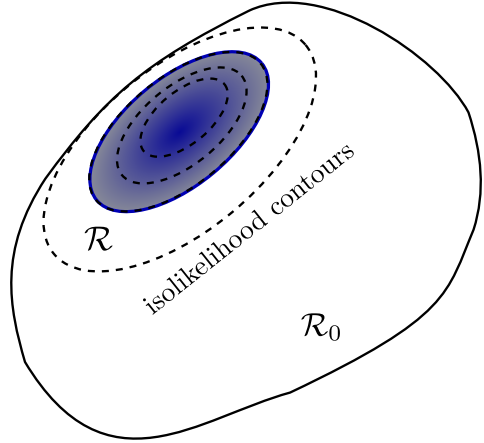}
	\caption{\label{fig:cred_reg} A credible region $\mathcal{R}=\mathcal{R}_\lambda$ (shaded) defined with some prior distribution $p(\rvec{r})$ in the parameter space $\mathcal{R}_0$ by the isolikelihood boundary of a $\lambda$ value.}
\end{figure}

The \emph{credible region} $\mathcal{R}$ for $\rvec{r}$ is the region of the smallest size for a fixed credibility, or equivalently the probability that $\rvec{r}$ is inside $\mathcal{R}$. In \cite{Shang:2013cc}, it was shown that $\mathcal{R}$ possesses an iso-likelihood boundary as illustrated in Fig.~\ref{fig:cred_reg}, which size and credibility are respectively
\begin{eqnarray}
s_\lambda&=&\,\int_{\mathcal{R}_0}(\D\,\rvec{r}')\,\chi_\lambda(\rvec{r}')\,,\,\,\chi_\lambda(\rvec{r})=\eta\left(L(\mathbb{D}|\rvec{r})-\lambda L_\mathrm{max}\right)\,,\nonumber\\
c_\lambda&=&\,\frac{1}{L(\mathbb{D})}\int_{\mathcal{R}_0}(\D\,\rvec{r}')\,\chi_\lambda(\rvec{r}')\,L(\mathbb{D}|\rvec{r}')\,.
\label{eq:sc_def}
\end{eqnarray}
The integration measure $(\D\,\rvec{r})$ should be understood as a product of the volume measure for the whole parameter space $\mathcal{R}_0$ \emph{and} the normalized prior probability distribution $p(\rvec{r})$ before the measurement is performed, which is part of the machinery in Bayesian statistics. Here $0\leq\lambda\leq1$ serves as the parameter that defines the likelihood $L(\mathbb{D}|\rvec{r})=\lambda L_\mathrm{max}$ in terms of its maximal value $L_\mathrm{max}=L(\mathbb{D}|\ML{\rvec{r}})$, and the region $\mathcal{R}=\mathcal{R}_\lambda\subset\mathcal{R}_0$ is specified by $\chi_\lambda(\rvec{r})$ [a Heaviside step function $\eta(\,\cdot\,)$] for the likelihood $L(\mathbb{D}|\rvec{r})$ describing the given physical situation. Thus, the credible region satisfies $L(\mathbb{D}|\rvec{r})/L_\mathrm{max}>\lambda$. We note here that minimizing size given a credibility is operationally dual to maximizing credibility for a given size and leads to the same optimal credible region.

There is also an important relationship between $s_\lambda$ and $c_\lambda$ that allows us to just compute $s_\lambda$ and infer $c_\lambda$ directly from it. This is written as~\cite{Shang:2013cc}
\begin{eqnarray}
c_\lambda=\dfrac{\displaystyle\lambda\,s_\lambda+\int^1_\lambda\D\lambda'\,s_{\lambda'}}{\displaystyle \int^1_0\D\lambda'\,s_{\lambda'}}\,.
\label{eq:cs_relation}
\end{eqnarray}

\subsubsection{Plausible regions}
\label{subsubsec:plaus_reg}
Inspecting $s_\lambda$ for a fixed $c_\lambda$, say 0.95, is a rather subjective choice. According to~\cite{Evans:2016aa}, we may exploit a statistically meaningful interpretation of the measured data to define another kind of Bayesian region. 

If we suppose that $\rvec{r}$ is \emph{plausibly} the true value, then we say that there is \emph{evidence} in favor of this supposition when its normalized posterior probability $L(\mathbb{D}|\rvec{r})\,p(\rvec{r})/L(\mathbb{D})$ $\left[L(\mathbb{D})=\int(\D\,\rvec{r}')\,L(\mathbb{D}|\rvec{r}')\leq L_\text{max}\right]$ is larger than its prior probability $p(\rvec{r})$. In other words, the evidence supports this prior knowledge. We can then construct another type of Bayesian region---the \emph{plausible region}---that contains \emph{all} plausible choices of $\rvec{r}$. This is the credible region $\mathcal{R}=\mathcal{R}_{\lambda=\lambda_{\rm{crit}}}$ characterized by the critical value~\cite{Li:2016da}
\begin{eqnarray}
\lambda_{\rm{crit}}=\int_0^1 d\lambda'\, s_{\lambda'}\,,
\label{eq:lbd_crit}
\end{eqnarray}
for which $L(\mathbb{D}|\rvec{r}\in\partial\mathcal{R}_{\lambda=\lambda_{\rm{crit}}})=L(\mathbb{D})$, or \emph{the credible region that contains all plausible points and nothing else}. To facilitate this understanding, we give a short instructive proof by noting that the constant $L(\mathbb{D})$ is simply related to the size function $s_\lambda$ by the definition
\begin{eqnarray}
L(\mathbb{D})&=&\,\int(\D\,\rvec{r}')\,L(\mathbb{D}|\rvec{r}')=\int(\D\,\rvec{r}')\,\int^{L(\mathbb{D}|\rvec{r}')}_0\D x'\nonumber\\
&=&\,L_\text{max}\int(\D\,\rvec{r}')\,\int^1_0\,\D\lambda'\,\eta\left(L(\mathbb{D}|\rvec{r}')-\lambda' L_\text{max}\right)=L_\text{max}\int_0^1 d\lambda'\, s_{\lambda'}\,,
\end{eqnarray}
so that the assignment $L(\mathbb{D}|\rvec{r}\in\partial\mathcal{R}_{\lambda=\lambda_{\rm{crit}}})\equiv\lambda_{\rm{crit}}L_\text{max}=L(\mathbb{D})$ gives the expression for $\lambda_{\rm{crit}}$.

\section{Analytical results for Bayesian regions}
\label{sec:results}

Throughout the discussions in this article, we shall assume that the parameter space $\mathcal{R}_0$ of $\rvec{r}$ is a convex space. The numerical computation of $s_\lambda$ and $c_\lambda$ for this convex space, and thereafter $\lambda_{\rm{crit}}$ for plausible regions, is known to be an NP-hard problem~\cite{Suess:2017np} because of the complicated influence from the parameter-space boundary $\partial\mathcal{R}_0$. In this section, we provide asymptotic analytical approximations for these quantities in the limit of large sample size $N\gg1$, which is the common regime in quantum estimation experiments. This allows an observer to make approximate error certification on $\ML{\rvec{r}}$ for any parameter dimension and measurements without performing intractable Monte Carlo calculations. In this limit, the likelihood $L(\mathbb{D}|\rvec{r})$ is approximately a Gaussian distribution. 

The choice of a \emph{prior} distribution $p(\rvec{r})$ for $\rvec{r}$ that makes up the integral measure $(\D\,\rvec{r})$ directly influences $s_\lambda$, which is the inherent nature of Bayesian analytics. For the purpose of revealing interesting properties of $s_\lambda$ and $c_\lambda$ through analytical expressions and avoid entangling with technical details of prior choices, we shall consider the uniform prior distribution over the parameters $\rvec{r}$, that is we take the \emph{primitive prior} $(\D\,\rvec{r})=\mathcal{N}\prod_j\D r_j$ of a suitable normalization constant $\mathcal{N}$.

We present results for three cases that can happen in quantum estimation. The first case is the rather optimistic scenario where the data $\mathbb{D}$ gives an estimator $\ML{\rvec{r}}$ that is well in the interior of $\mathcal{R}_0$, such that the Gaussian likelihood is mainly contained in $\mathcal{R}_0$. The second case, which happens much more frequently when the true parameter $\rvec{r}$ is exactly in $\partial\mathcal{R}_0$, describes an interior-point $\ML{\rvec{r}}$ that is near the boundary $\partial\mathcal{R}_0$ with the Gaussian likelihood partially truncated. The third case, which is again prevalent if $\rvec{r}\in\partial\mathcal{R}_0$, is where the ML estimator $\ML{\rvec{r}}$ lies exactly on $\partial\mathcal{R}_0$. Closed-form expressions for $s_\lambda$, $c_\lambda$ and $\lambda_{\rm{crit}}$ are easily obtainable for the first case, whereas for the second and third cases, concise analytical approximations are available only for $s_\lambda$, from which $c_\lambda$ and $\lambda_{\rm{crit}}$ can be tractably inferred using the respective simple relations in \eqref{eq:cs_relation} and \eqref{eq:lbd_crit}. However for single-parameter estimation settings, exact analytical expressions for the second and third cases are available.

\subsection{Case 1: Interior-point theory for a full likelihood}
For a $d$-dimensional parameter $\rvec{r}$, if $\rvec{r}\notin\partial\mathcal{R}_0$, then for a given data $\mathbb{D}$ collected with sufficiently large number of copies $N$, we approximate the likelihood
\begin{eqnarray}
L(\mathbb{D}|\rvec{r})\approx L_\mathrm{max}\,\exp\left(-\frac{1}{2}\rvec{\Delta}(\rvec{r})\bcdot\FML\bcdot\rvec{\Delta}(\rvec{r})\right),
\label{eq:gauss_approx}
\end{eqnarray}
with a Gaussian function~\cite{Rehacek:2008ml} centered at the experimentally-obtained $\ML{\rvec{r}}$ that has a covariance equal to the $d$-dimensional Fisher information\footnote{A prudent observer might consider the negative of the Hessian  $\dyadic{H}(\rvec{r})=\displaystyle\sum_k\dfrac{n_k}{p_k}\left(-\dfrac{1}{p_k}\dfrac{\partial p_k}{\partial\rvec{r}}+\dfrac{\partial}{\partial\rvec{r}}\right)\dfrac{\partial p_k}{\partial\rvec{r}}$ for finite $N$ instead of the Fisher information.}
\begin{eqnarray}
\dyadic{F}(\rvec{r})&=&\,\sum_k\frac{N}{p_k}\frac{\partial p_k}{\partial \rvec{r}}\frac{\partial p_k}{\partial\rvec{r}}\,,\nonumber\\
\rvec{\Delta}(\rvec{r})&=&\,\rvec{r}-\ML{\rvec{r}}
\end{eqnarray}
evaluated at $\ML{\rvec{r}}$---$\FML=\dyadic{F}(\ML{\rvec{r}})$ for multinomial data statistics.

\begin{figure}[t]
	\center
	\includegraphics[width=0.7\columnwidth]{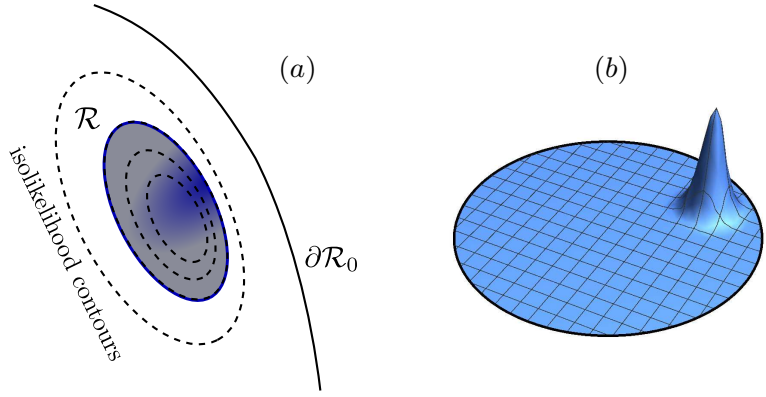}
	\caption{\label{fig:Fig_int1} (a)~The Bayesian region $\mathcal{R}$ is centered at an interior ML estimator $\ML{\rvec{r}}$, such that (b)~the width of the likelihood function (blue plot bounded by the convex boundary $\partial\mathcal{R}_0$) is mainly contained within the parameter space unless $\lambda$ is extremely small. The truncated tails of the likelihood by $\partial\mathcal{R}_0$ give no statistical contribution to the Bayesian region as long as $N$ is sufficiently large. If the volume of $\mathcal{R}_0$ is large, this condition is usually achievable without a very large $N$.}
\end{figure}

\begin{figure}[t]
	\center
	\includegraphics[width=0.8\columnwidth]{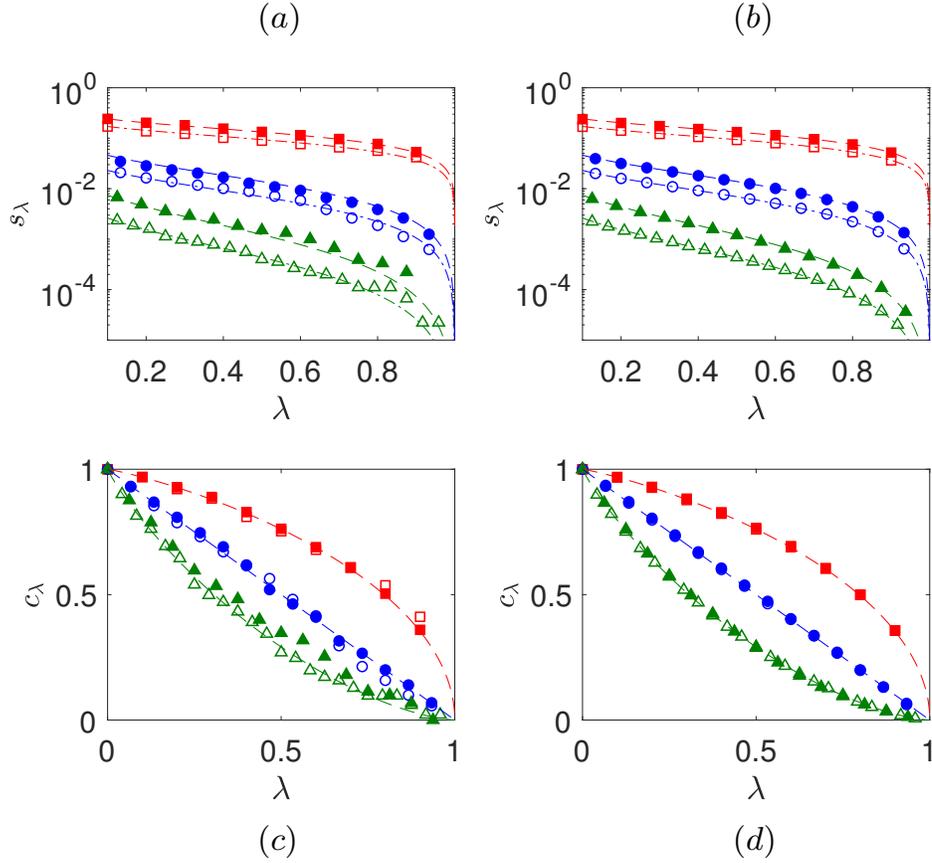}
	\caption{\label{fig:s_c_behav_1} Characteristic plots of (a,b) $s_\lambda$ (logarithmic) and (c,d) $c_\lambda$ (linear) against $\lambda$ for Gaussian distributions. Panels (a) and (c) refer to one numerical Gaussian sampling experiment, whereas panels (b) and (d) refer to an average over 100 experiments. The square, circular and triangular markers plot data for one, two and three-dimensional Gaussian distributions, each of which is specified by a randomly-chosen covariance (a random positive matrix). Filled markers correspond to $N=100$ while the unfilled ones correspond to $N=500$. The dashed curves represent analytical values of Eq.~\eqref{eq:interior_cred}. We note from the plots that for larger $N$, accurate computations of $s_\lambda$ and $c_\lambda$ require very large numbers of $\lambda$ divisions for the numerical integrations, which we cap at a certain number. An average over experiments seem to reduce the fluctuations from inaccurate numerical integrations with finite numbers of $\lambda$ values.}
\end{figure}

As mentioned in the caption of Fig.~\ref{fig:Fig_int1}, if the (prior-influenced) volume $V_{\mathcal{R}_0}$ of $\mathcal{R}_0$ is large enough, then typically an interior ML estimator can be obtained with no likelihood truncation without a very large $N$. This applies to the estimation of one or few interferometer phases, tomography of a single qubit, \emph{etc}, where the volume of $\mathcal{R}_0$ is not restricted by too many parameter convex constraints. Under this condition, it is easy to see that $\mathcal{R}$ is a full hyperellipsoid which volume is defined by $\lambda$ and the prior $p(\rvec{r})$. For the uniform prior, we may either take well-known statements in, say, \cite{Albert:2016ag} or simply work out the expressions from \eqref{eq:sc_def} as in \ref{app:cs_int1}. Either way, we have the interior-point expressions
\begin{eqnarray}
s_\lambda&=&\,\frac{V_d}{V_{\mathcal{R}_0}}(-2\log\lambda)^{d/2}\,\DET{\FML}^{-1/2}\,,\nonumber\\
c_\lambda&=&\,1-\frac{\Gamma(d/2, -\log\lambda)}{(d/2-1)!}\,,
\label{eq:interior_cred}
\end{eqnarray}
where $V_d=\pi^{d/2}/(d/2)!$ is the volume of the $(d-1)$-sphere of unit radius, and $\Gamma(a,y)$ is the order-$a$ upper incomplete Gamma function of $y$. We may also express $s_\lambda=V_{d,\lambda}/V_{\mathcal{R}_0}$ in terms the normalized hyperellipsoidal volume $V_{d,\lambda}=V_d(-2\log\lambda)^{d/2}\,\DET{\FML}^{-1/2}$.

\begin{figure}[t]
	\center
	\includegraphics[width=0.8\columnwidth]{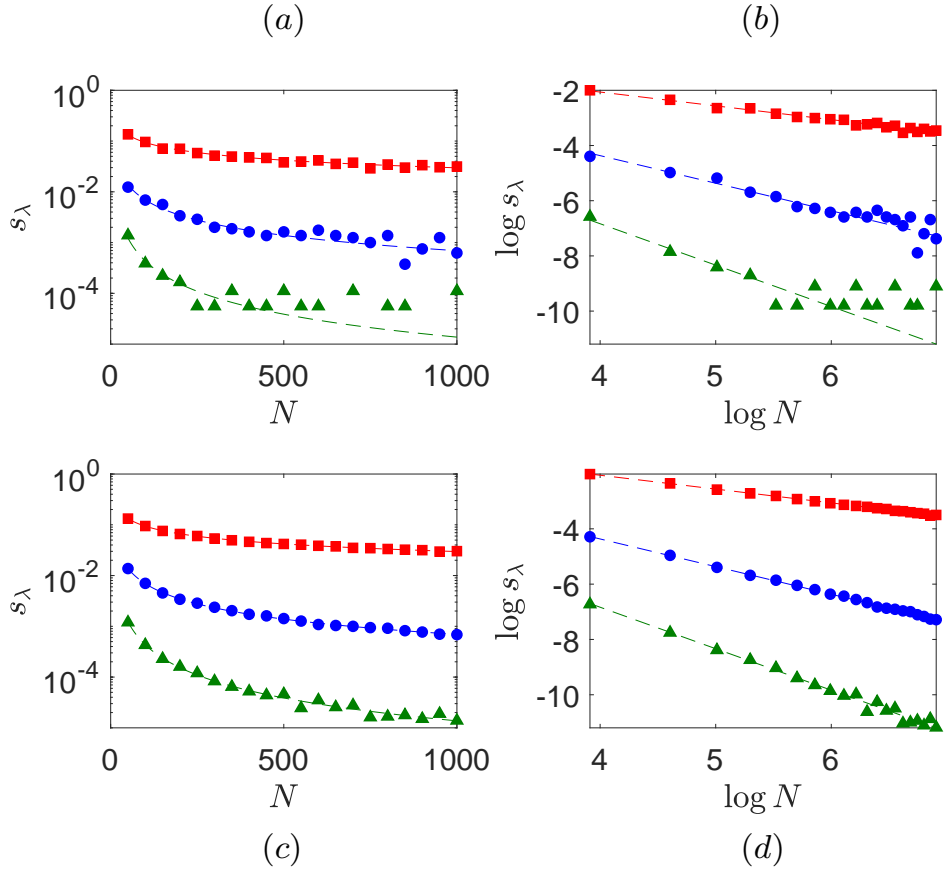}
	\caption{\label{fig:s_c_behav_2} Characteristic plots of $s_\lambda$ (95\%-credible regions) against $N$ in (a,c) linear and (b,d) logarithmic scales. The two sets of graphs in (a) and (b) refer to one particular experiment, and those in (c) and (d) refer to an average over many experiments. Data marker descriptions follow Fig.~\ref{fig:s_c_behav_1}.}
\end{figure}

In this optimistic case, the size $s_\lambda$ converges \emph{logarithmically} in $\lambda$. Furthermore, the simple form of $c_\lambda$ allows us to express $s_\lambda$ as a function of $c_\lambda$, namely
\begin{eqnarray}
s_\lambda=\frac{V_d}{V_{\mathcal{R}_0}}\left[2\,\Gamma^{-1}_{d/2}\left(1-c_\lambda\right)\right]^{d/2}\,\DET{\FML}^{-1/2}\,,
\end{eqnarray}
where the inverse $\Gamma^{-1}_{a}(y)$ of the \emph{regularized} incomplete Gamma function can be numerically computed efficiently~\cite{Didonato:1986aa}. 

Since we have assumed that each measurement copy is i.i.d., the Fisher information $\FML$ is proportional to $N$. It follows straightforwardly that in the large-$N$ limit, the size $s_\lambda$ scales according to $1/N^{d/2}$ (that is a contribution of $1/\sqrt{N}$ for every dimension), whereas the credibility $c_\lambda$ is independent of $N$. These scaling behaviors can be observed in Figs.~\ref{fig:s_c_behav_1} and \ref{fig:s_c_behav_2}, where the important characteristics of these two region quantites are tested in mean-estimation simulations for Gaussian distributions of various dimensions $d$ and given covariances. 

Under the Gaussian approximation in \eref{eq:gauss_approx}, we can easily obtain
\begin{eqnarray}
	\lambda_{\rm{crit}}=\sqrt{\DET{2\pi\,\FML^{-1}}}/V_{\mathcal{R}_0}\,,
	\label{eq:lbd_crit_case1}
\end{eqnarray}
and so the plausible region possesses a size and credibility given by
\begin{eqnarray}
s_{\lambda_{\rm{crit}}}&=&\,\frac{V_d}{V_{\mathcal{R}_0}}\left[-\log\left(\dfrac{\DET{2\pi\,\FML^{-1}}}{V^2_{\mathcal{R}_0}}\right)\right]^{d/2}\,\DET{\FML}^{-1/2}\,,\nonumber\\
c_{\lambda_{\rm{crit}}}&\approx&\,1-\frac{(d/2)^{d/2-1}}{(d/2-1)!}\frac{(\log N)^{d/2-1}}{N^{d/2}}\,,
\label{eq:interior_plaus}
\end{eqnarray}
We note here that for the plausible region, the scaling behaviors of $s_{\lambda_\text{crit}}$ and $c_{\lambda_\text{crit}}$ with $N$ are more complicated. For i.i.d. copies, we have $s_{\lambda_\text{crit}}\sim(\log N+\cdots)^{d/2}/N^{d/2}$ and $1-c_{\lambda_\text{crit}}\sim(\log N)^{d/2-1}/N^{d/2}$, where the appearance of logarithmic scaling comes from picking the largest credible region that contains all plausible parameters~(explained in Sec.~\ref{subsubsec:plaus_reg}).

\begin{figure}[t]
	\center
	\includegraphics[width=0.9\columnwidth]{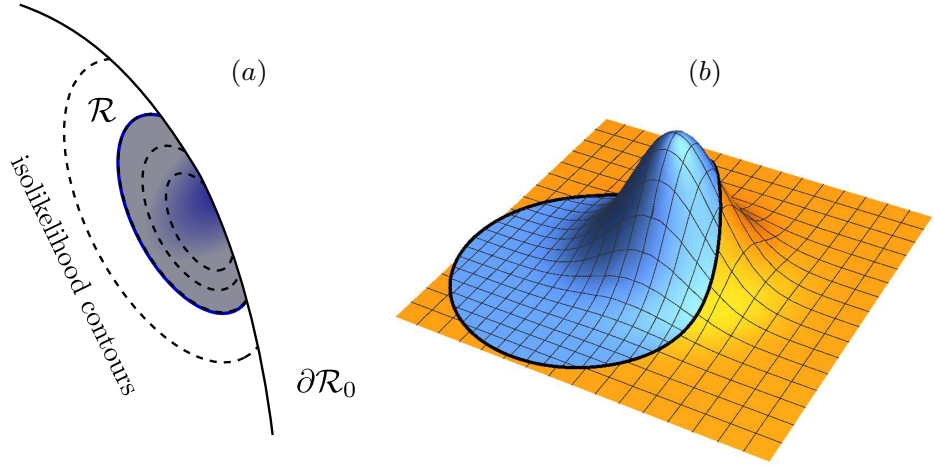}
	\caption{\label{fig:Fig_int2} (a)~The Bayesian region $\mathcal{R}$ is centered at an interior $\ML{\rvec{r}}$ that is quite close to the boundary $\partial\mathcal{R}_0$, resulting in (b)~the truncation of a significant portion of the likelihood (orange surface covers points outside of $\partial\mathcal{R}_0$). This occurs quite often whenever $\rvec{r}\in\partial\mathcal{R}_0$ and $N$ is not large enough to avoid the influence of $\partial\mathcal{R}_0$ even though the Gaussian approximation in \eqref{eq:gauss_approx} is accurate.}
\end{figure}

\begin{figure}[t]
	\center
	\includegraphics[width=0.65\columnwidth]{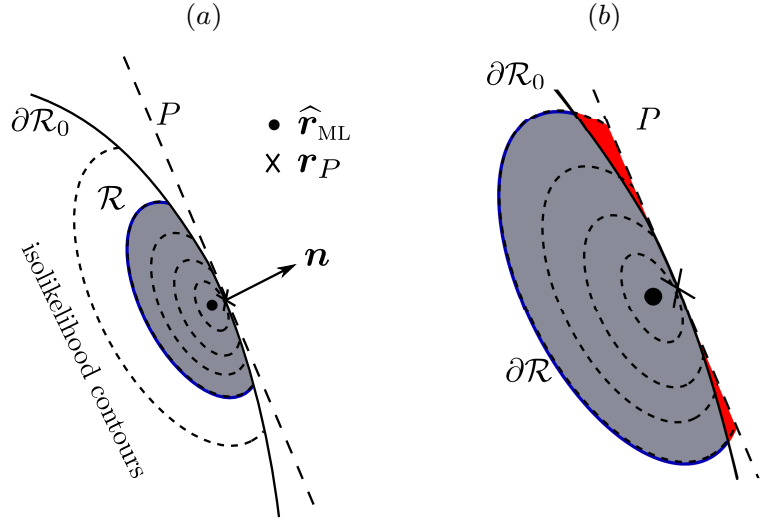}
	\caption{\label{fig:R0_hyperplane}(a)~If $\lambda$ is small enough so that the Bayesian region $\mathcal{R}$ is truncated, then approximating the joint boundary $\partial\mathcal{R}\cap\partial\mathcal{R}_0$ with a hyperplane $P$ allows us to estimate the volume of the actual truncated hyperellipsoid $\mathcal{R}$. (b)~The discrepancy (red shaded region bounded by $\partial\mathcal{R}\cap\partial\mathcal{R}_0$ and $P$) asymptotically goes to zero as $N$ increases when $\partial\mathcal{R}\cap\partial\mathcal{R}_0$ is smooth.}
\end{figure}

\subsection{Case~2: Interior-point theory for a truncated likelihood}

For the case of an interior ML estimator, the more frequent case would be that part of $\mathcal{R}$ is truncated by the boundary of the convex parameter space $\mathcal{R}_0$~(see Fig.~\ref{fig:Fig_int2}). This can occur when $N$ is not large enough to shrink the uncertainty of the estimator so that $\mathcal{R}$ is completely interior, especially when the true parameter $\rvec{r}$ lies on the boundary $\partial\mathcal{R}_0$. The geometry of $\mathcal{R}=\mathcal{R}_\lambda$ for interesting values of $\lambda$ is now a \emph{truncated hyperellipsoid} of center $\ML{\rvec{r}}$, and the boundary effect of the parameter space cannot be neglected in this case. Nonetheless, the problem of calculating $s_\lambda$ and $c_\lambda$ is now equivalent to finding the fraction of the hyperellipsoidal volume [dictated by \eqref{eq:interior_cred}] that is removed by $\partial\mathcal{R}_0$. 

Solving this problem requires the identification of the boundary for $\mathcal{R}$, which is computationally hard. We therefore investigate the limit when $N$ is sufficiently large enough so that the joint boundary $\partial\mathcal{R}\cap\partial\mathcal{R}_0$, as depicted in Fig.~\ref{fig:R0_hyperplane}, (i) has no disjointed regions and (ii) is approximately a hyperplane $P$ containing the boundary point $\rvec{r}_P$ with the largest likelihood. This hyperplane $P$ has a normal $\rvec{n}$ that is orthogonal to the isolikelihood contour at $\rvec{r}_P$. As $\partial\mathcal{R}_0$ is not convex, maximizing the likelihood over $\partial\mathcal{R}_0$ is typically a difficult problem and one always has to rely on heuristic numerical methods. On the other hand, since $\ML{\rvec{r}}$ is near $\partial\mathcal{R}\cap\partial\mathcal{R}_0$, clearly $\norm{\rvec{r}_P-\ML{\rvec{r}}}$ is small and we may exploit this fact to estimate $\rvec{r}_P$, and the corresponding maximal likelihood value $L^{(\partial\mathcal{R}_0)}_\text{max}$ with a simple Monte Carlo algorithm in \ref{app:case2_rP}. 

After obtaining $L^{(\partial\mathcal{R}_0)}_\text{max}=L_\text{max}\exp\left(-\rvec{\Delta}(\rvec{r}_P)\bm{\cdot}\FML\bm{\cdot}\rvec{\Delta}(\rvec{r}_P)/2\right)$, it is possible to show that the estimated fraction $\gamma$ of the hyperellipsoid truncation is given in terms of the regularized incomplete beta function $\BETAr{y}{a,b}$ as
\begin{eqnarray}
\gamma&=&\,1-\BETAr{\frac{1-l}{2}}{\frac{d+1}{2},\frac{d+1}{2}}\,,\nonumber\\
l&=&\,\min\left\{\sqrt{\dfrac{\log \lambda_\text{int}}{\log\lambda}},1\right\}\,,
\lambda_\text{int}=\dfrac{L^{(\partial\mathcal{R}_0)}_\text{max}}{L_\text{max}}\,,
\label{eq:case2_1}
\end{eqnarray}
with which we arrive at the generalized interior-point statement $s_\lambda\approx\gamma V_{d,\lambda}/V_{\mathcal{R}_0}$. For $\lambda=\lambda_\text{int}$, $\gamma=1$ characterizes the optimistic size expression in \eqref{eq:interior_cred}. The approximate credibility has no simple closed form but may be computed with the relation in \eqref{eq:cs_relation} efficiently. 

Details of the derivation of \eqref{eq:case2_1} is given in \ref{app:case2_deriv}. More relevantly, Let us briefly discuss the volume estimate characterized by the fraction in \eqref{eq:case2_1} in broad terms. For this, we emphasize that $\partial\mathcal{R}_0$ can be a highly sophisticated surface with corners and edges. For instance, if $\mathcal{R}_0$ is the space of quantum states of Hilbert-space dimension $D=2$---the qubit space---, then $\partial\mathcal{R}_0$ that is enforced by the operator positivity constraint is a 2-sphere. However if $D>2$, $\partial\mathcal{R}_0$ is generally a complicated surface with corners and edges, for the convex space is ``neither a polytope nor a smooth body.''~\cite{Bengtsson:2013gm} For such boundaries, the approximated volume fraction offered by \eqref{eq:case2_1} is an overestimate of the actual fraction for any finite $N$ due to the convex nature of $\mathcal{R}_0$. If however $\ML{\rvec{r}}$ lies on a smooth $\partial\mathcal{R}\cap\partial\mathcal{R}_0$ to which we may approximate the local boundary with a hyperplane, then in the limit of large $N$, this overestimate approaches the exact answer, which applies, for instance, to the qubit space.

It is easy to see that this methodology gives the asymptotically exact, not an overestimated volume fraction in single-parameter estimation ($d=1$), as the $\partial\mathcal{R}\cap\partial\mathcal{R}_0$ intersects $P$ at exactly the point $r_P$. We note, however, that the likelihood near $r_P$ is exponential in $r$. The corresponding quantities $s_\lambda$, $c_\lambda$ and $\lambda_\text{crit}$ also admit analytical expressions
\begin{eqnarray}
s_\lambda&=&\,V_{1,\lambda}+\eta(\lambda_\text{int}-\lambda)\dfrac{\log\lambda-\log\lambda_\text{int}}{V_{\mathcal{R}_0}|\Delta(r_P)|}\,,\nonumber\\
c_\lambda&=&\,\dfrac{|\Delta(r_P)|\sqrt{2F_\textsc{ml}}\left[\sqrt{\pi}-\Gamma\left(1/2,-\log\lambda\right)\right]+\eta(\lambda_\text{int}-\lambda)(\lambda-\lambda_\text{int})}{\sqrt{2\pi F_\textsc{ml}}|\Delta(r_P)|-\lambda_\text{int}}\,,\nonumber\\
\lambda_\text{crit}&=&\,\dfrac{\sqrt{2\pi}}{V_{\mathcal{R}_0}\sqrt{F_\textsc{ml}}}-\dfrac{\lambda_\text{int}}{V_{\mathcal{R}_0}|\Delta(r_P)|}\,,
\label{eq:case2_2}
\end{eqnarray}
which can be derived by evaluating the one-dimensional version of the integral in \eqref{eq:Vintersect}. The limiting case in which $\lambda_\text{int}\rightarrow0$ can be confirmed right away.

\begin{figure}[t]
	\center
	\includegraphics[width=0.8\columnwidth]{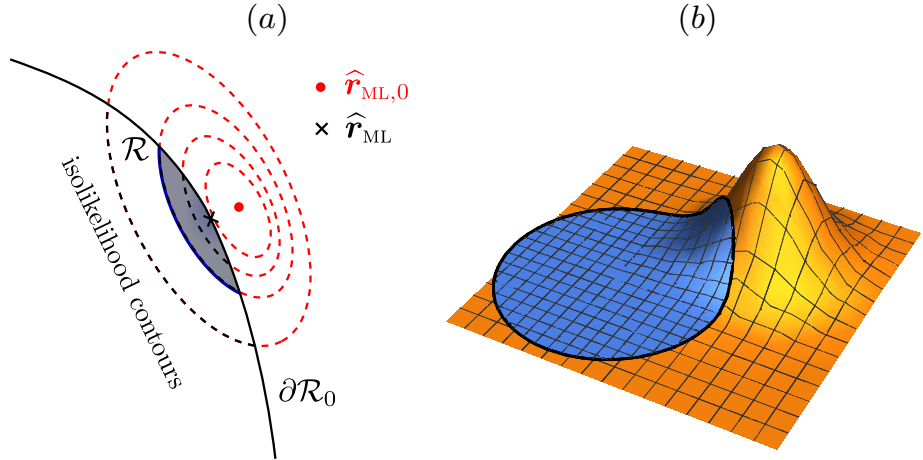}
	\caption{\label{fig:bd_1}The case where $\ML{\rvec{r}}$ lies precisely on the boundary $\mathcal{R}\cap\mathcal{R}_0$ is predominantly due to the fact that (a)~the actual ML estimator $\MLo{\rvec{r}}\notin\mathcal{R}_0$. That $\ML{\rvec{r}}=\MLo{\rvec{r}}$ in this case is a measure-zero event. Such an observation is routine for a boundary-point $\rvec{r}$. (b)~The corresponding likelihood (orange) peak that is outside of $\mathcal{R}_0$ gives the maximum achievable value if the convex boundary $\partial\mathcal{R}_0$ is relaxed. Otherwise, the maximum of the likelihood function over $\mathcal{R}_0$ would be $\ML{\rvec{r}}$.}
\end{figure}

\subsection{Case~3: Boundary-point theory}

If $\ML{\rvec{r}}$ is on the joint boundary $\partial\mathcal{R}\cap\partial\mathcal{R}_0$, this practically means that if one actively searches for the ML estimator without the external constraints of the parameters, the maximum $\MLo{\rvec{r}}$ that corresponds to this search will lie outside of $\mathcal{R}_0$~(see Fig.~\ref{fig:bd_1}). The single-phase estimation of a Mach-Zehnder interferometer is a simple one-dimensional example where if the unknown true phase $0\leq\theta\leq2\pi$ is restricted in the interval $\theta_a\leq\theta\leq\theta_b$ (possibly by some prior or physical limitations), then there can be a situation in which the $\ML{\theta}\neq\MLo{\theta}$, or equivalently $\ML{\theta}=\theta_a$ or $\theta_b$. Another important example is state tomography where if the true state $\rvec{r}\rightarrow\rho$ is on the boundary of the state space, then there is a high probability that $\MLo{\rho}$ lies outside the space and $\ML{\rho}$ is a rank-deficient estimator.

\begin{figure}[t]
	\center
	\includegraphics[width=0.8\columnwidth]{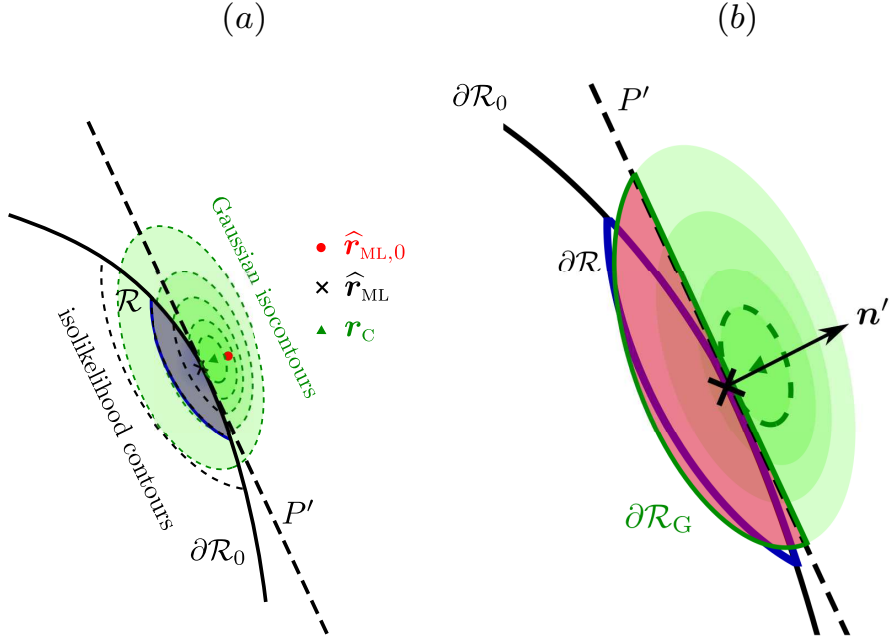}
	\caption{\label{fig:bd_2}In coping with the boundary-point case, (a)~an expansion of the likelihood about the correct $\ML{\rvec{r}}$ to second order in $\rvec{r}-\ML{\rvec{r}}$ gives a new Gaussian approximation (green) that is centered at $\rvec{r}_\text{c}$. If $N$ is large enough, the Gaussian isocontours will match the isolikelihood contours closely---$\rvec{r}_\textsc{c}\approx\MLo{\rvec{r}}\approx\ML{\rvec{r}}$. (b)~The corresponding estimate for $\mathcal{R}$ is then the region (pink shaded) bounded by the Gaussian isocontour for $\lambda$ and the hyperplane $P'$ that contains $\ML{\rvec{r}}$ and has a normal $\rvec{n}'$ perpendicular to the isocontour intersecting $\ML{\rvec{r}}$. One may then estimate $V_{\mathcal{R}}$ by the volume of this region. For smooth $\partial\mathcal{R}\cap\partial\mathcal{R}_0$ and sufficiently large $N$, this estimate is asymptotically exact.}
\end{figure}

With the statistical conviction that the true parameter $\rvec{r}$ is close to the boundary-point ML estimator $\ML{\rvec{r}}$, we may again expand $\log L(\mathbb{D}|\rvec{r})$ to second order,
\begin{eqnarray}
\log L(\mathbb{D}|\rvec{r})&\approx&\,\log L_{\text{max}}+\rvec{\Delta}(\rvec{r})\bm{\cdot}\gML-\frac{1}{2}\,\rvec{\Delta}(\rvec{r})\bm{\cdot}\FML\bm{\cdot}\rvec{\Delta}(\rvec{r})\,,\nonumber\\
\qquad\quad\gML&=&\,\rvec{\partial}_\textsc{ml}\log L(\mathbb{D}|\ML{\rvec{r}})\,,
\label{eq:gauss_approx2}
\end{eqnarray}
where now evidently the first order does not vanish since $\ML{\rvec{r}}$ is on the boundary and $L_\text{max}$, the maximal likelihood value for $\mathcal{R}_0$, is less than the exterior maximal value $L_\text{max,G}=L_\text{max}\exp\left(\gML\bm{\cdot}\FML^{-1}\bm{\cdot}\gML/2\right)$ for the approximated Gaussian function. Similar to Case~2, we may introduce a hyperplane $P'$ that contains $\ML{\rvec{r}}$ and has a normal $\rvec{n}'=\gML$ that is orthogonal to the Gaussian isocontour intersecting $\ML{\rvec{r}}$. The volume $V_{\mathcal{R}}$ of $\mathcal{R}$ can then be (over)estimated with the shaded volume presented in Fig.~\ref{fig:bd_2}. For smooth boundaries, this estimate once more becomes asymptotically exact.

Interestingly, we point out the role changes for some relevant quantities: $\ML{\rvec{r}}$ now takes the place of $\rvec{r}_P$ as the boundary point in the hyperplane and $L_\text{max,G}$ is now replacing $L_\text{max}$ to be the largest possible likelihood. We may next define $\lambda_{\rm{eff}}=\lambda L_\text{max}/L_\text{max,G}<1$ to be the effective ``$\lambda$'' that characterizes the approximated Gaussian likelihood with respect to the actual one. Finally, after realizing that the estimated volume for $V_{\mathcal{R}}$ falls on the opposite side of the hyperplane in contrast with that in Case~2, we can write down the fraction
\begin{eqnarray}
\gamma'=\BETAr{\frac{1-l'}{2}}{\frac{d+1}{2},\frac{d+1}{2}}\,,\quad l'=\sqrt{\dfrac{\log \lambda_\text{bd}}{\log\lambda_\text{eff}}}\leq1\,,\quad \lambda_\text{bd}=\dfrac{L_\text{max}}{L_\text{max,G}}\,,
\label{eq:case3_1}
\end{eqnarray}
of the total hyperellipsoidal volume that contributes to the approximate size estimate $s_\lambda\approx\gamma' V_{d,\lambda}/V_{\mathcal{R}_0}$.

The asymptotically exact region quantities for $d=1$ can be obtained by taking the aforementioned role changes into account. This suggests the replacements in \eqref{eq:case2_2} (\emph{from Case~2 to Case~3}) $|\Delta(r_P)|\rightarrow g_\textsc{ml}$, $\lambda_{\rm{int}}\rightarrow\lambda_\text{eff}/\lambda$ and $\lambda\rightarrow\lambda_{\rm{eff}}$, which immediately gives rise to
\begin{eqnarray}
s_\lambda=-\frac{\log\lambda}{V_{\mathcal{R}_0}\,g_{\textsc{ml}}}\,,\quad
\quad c_\lambda=1-\lambda\,,\quad\lambda_{\rm{crit}}=\dfrac{1}{V_{\mathcal{R}_0}\,g_{\textsc{ml}}}\,,
\label{eq:case3_2}
\end{eqnarray}
with the appropriate sign changes due to the opposite ``side'' of the truncation to Case~2.

\subsection{Remarks on logarithmic divergence and $V_{\mathcal{R}_0}$}

In all the Bayesian-region property formulas developed [\eqref{eq:interior_cred}, \eqref{eq:interior_plaus}, \eqref{eq:case2_1}, \eqref{eq:case2_2}, \eqref{eq:case3_1}, \eqref{eq:case3_2}] as a means to provide an asymptotic size and credibility certification for the ML estimator $\ML{\rvec{r}}$, the size formulas exhibit logarithmic divergences---$s_\lambda\sim(-\log\lambda)^{d/2}$. This feature stems from the Gaussian approximations in \eqref{eq:gauss_approx} and \eqref{eq:gauss_approx2} that pays no attention to the parameter-space boundary $\partial\mathcal{R}_0 \backslash(\partial\mathcal{R}\cap\partial\mathcal{R}_0)$ that falls on ``the other side'' of the joint one (if there is any). These approximations are strictly valid for the likelihood portion sufficiently near the maximum. For extremely small $\lambda$ values or high credibilities, the asymptotic size formulas either give highly conservative (much larger) estimates for $s_\lambda$, or gradually exceeds the unit physical upper bound.

This reinforces the importance of measuring a sufficiently large number of copies $N$ such that most portion of the likelihood is approximately part of a Gaussian function. Put differently, there exists the sufficient condition 
\begin{eqnarray}
N\gg N_\text{min}\quad\text{where}\quad\left.\rvec{\Delta}(\rvec{r}_P)\bm{\cdot}\FML\bm{\cdot}\rvec{\Delta}(\rvec{r}_P)\right|_{N=N_\text{min}}=-2\log\lambda
\end{eqnarray}
given a particularly interesting range of $\lambda$. This is geometrically equivalent to keeping the tails of the likelihood from penetrating the boundary $\partial\mathcal{R}_0\neq\partial\mathcal{R}\cap\partial\mathcal{R}_0$ as much as possible, so that the logarithmic divergence has no visible effect on the size estimation.

Furthermore, all operational formulas invoke the knowledge of the volume $V_{\mathcal{R}_0}$ of $\mathcal{R}_0$ under the uniform-prior assertion. For parameter estimation settings with simple convex boundary constraints this can be found very easily. For instance, $V_{\mathcal{R}_0}$ for an \emph{a priori} uniformly distributed phase $a\leq\theta\leq b$ is $b-a$. In the case of quantum-state characterization $V_{\mathcal{R}_0}$ is much more complicated, but known to have closed forms for specialized priors~\cite{Zyczkowski:2003qs,Andai:2006qs}. Just as an example we shall take the prior to be the uniform distribution over the continuous space $\mathcal{R}_0=\mathcal{M}_D$ of $D$-dimensional complex positive matrices of unit trace that represent quantum states $\rho$, or the \emph{Lebesgue prior} for this space. For this prior, the volume for the $(d=D^2-1)$-dimensional state parameter $\rvec{r}$ has the closed form~\cite{Andai:2006qs}
\begin{eqnarray}
V_{\mathcal{M}_D}=\dfrac{\pi^{D(D-1)/2}}{(D^2-1)}\prod^{D-1}_{j=1}j!\,.
\label{eq:leb_vol}
\end{eqnarray}

\begin{figure}[t]
	\center
	\includegraphics[width=1\columnwidth]{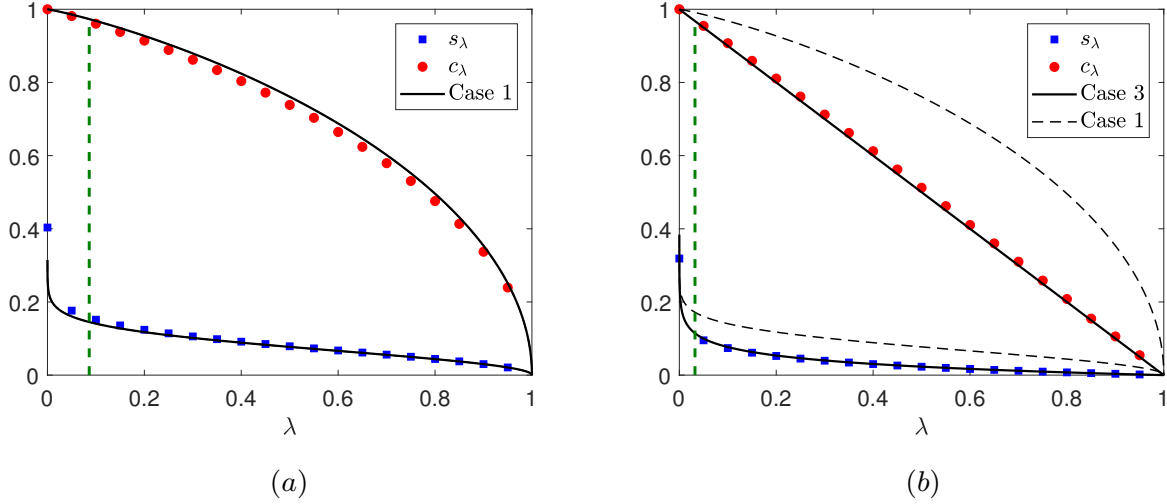}
	\caption{\label{fig:cred_1D}Single-parameter qubit estimation. (a)~For a one-dimensional qubit in a mixed state specified by $r=0.99$, $N=30$ is sufficiently large for boundary effects of $\mathcal{M}_2$ to vanish, which explains the accuracy of the interior-point expressions in~\eqref{eq:interior_cred}. The plausible region, of 0.966 credibility, is defined with $\lambda_\mathrm{crit}=0.08$ (dashed line). (b)~In the case where $\ML{r}=1$ is in $\partial\mathcal{R}\cap\partial\mathcal{R}_0$, while $N=30$ avoids the tail-boundary effects at $r=0$, the part at $r=1$ modifies the behaviors of $s_\lambda$ and $c_\lambda$ according to~\eqref{eq:case3_2}. Here, the plausible region, of 0.967 credibility, is constructed with $\lambda_\mathrm{crit}=0.03$.}
\end{figure}

\section{Examples in quantum-state tomography}
\label{sec:examples}

\subsection{Qubit}
In quantum-state tomography of a single-qubit $(D=2)$, the space $\mathcal{R}_0=\mathcal{M}_2$ of statistical operators can be conveniently represented as the $2\times2$ complex positive matrix
\begin{eqnarray}
\rho\,\,\widehat{=}\begin{pmatrix}
r_1 & r_2-\I r_3\\
r_2+\I r_3&1-r_1
\end{pmatrix}
\label{eq:qubit_par}
\end{eqnarray}
in terms of the $(d=3)$-dimensional state parameter $\rvec{r}$. The qubit space also has the nice property that the boundary $\partial\mathcal{M}_2$ is smooth---it is the surface of a 2-sphere. This implies that $\partial\mathcal{M}_2$ is smooth and can eventually be described by a hyperplane for sufficiently large $N$. We shall see that the expressions in \eqref{eq:case2_1}, \eqref{eq:case2_2}, \eqref{eq:case3_1} and \eqref{eq:case3_2} indeed exactly describe the actual size and credibility in this limit. 

To verify our theoretical results, we may consider three different classes of qubit states. For the numerical computation of $s_\lambda$ and $c_\lambda$, one may first generate a set of qubit states for the integrations by performing uniform rejection sampling. In accordance with the Lebesgue measure, the parametrization in \eqref{eq:qubit_par} allows a uniform sampling on the parameter ranges $0\leq r_1\leq1$ and $-1\leq r_2,r_3\leq 1$ depending on the class of qubit states, where the range of $r_1$ trivially maintains the unit-trace constraint. From this set of random operators,rejection sampling is then carried out by simply eliminating randomly generated operators this way that are not positive. These matrices may be numerically filtered out by verifying efficiently that their Cholesky decompositions do not exist~\cite{Rump:2006aa}. In what follows, the yield percentage from uniform rejection sampling, that is the percentage ratio of the number of positive operators out of the total number of sampled Hermitian operators, is calculated explicitly for each of the three classes.

\subsubsection{One-parameter qubit $(d=1)$}
Suppose we know that $\rho$ corresponds to $r_2=r_3=0$, so that only the single parameter $r=r_1$ needs to be estimated. The POM considered shall then be the simple ($M=2$)-outcome projective measurement onto the eigenstates of $\sigma_z=\ket{0}\bra{0}-\ket{1}\bra{1}$ that directly probes $r$,
\begin{eqnarray}
p_1&=\opinner{0}{\rho}{0}=r\,,\nonumber\\
p_2&=\opinner{1}{\rho}{1}=1-r\,.
\end{eqnarray}

\begin{figure}[t]
	\center
	\includegraphics[width=1\columnwidth]{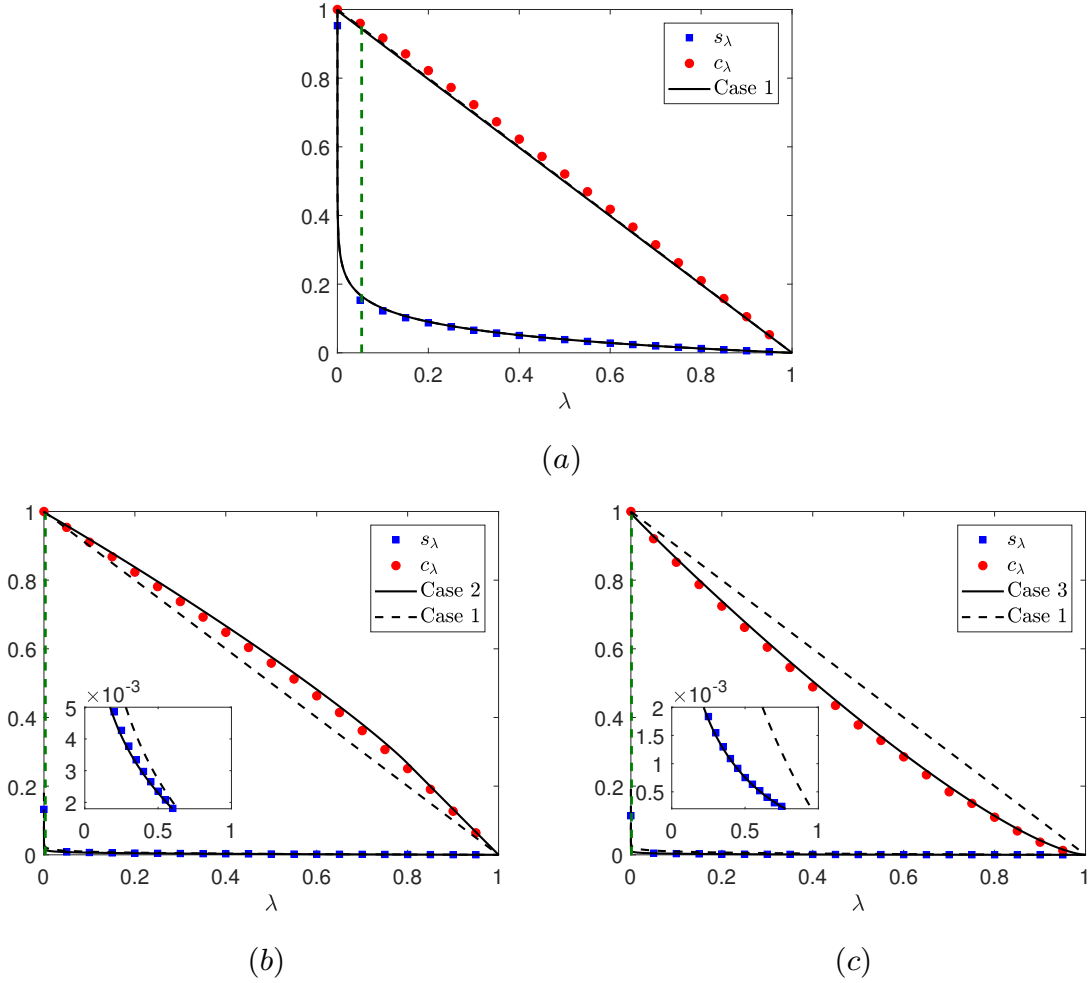}
	\caption{\label{fig:cred_2D}Two-parameter qubit estimation. (a)~Tomography is carried out on a two-dimensional qubit which quantum state is represented by $\rvec{r}=\TP{(0.8\,\,0.1)}$ inside the Bloch ball. The interior ML estimator $\ML{\rvec{r}}$ for $N=50$ is far enough from the boundary so that the results of Case~1 apply. The plausible region of 0.957 credibility is defined by $\lambda_\mathrm{crit}\approx0.05$. (b)~For a different state $\rvec{r}=\TP{(0.8\,\,0.4)}$, $\ML{\rvec{r}}$ for $N=500$ is near $\partial\mathcal{R}\cap\partial\mathcal{R}_0$ and the generalized solutions for Case~2 clearly resolve the curvature modifications on $s_\lambda$ (see also the inset for a blown up plot of $s_\lambda$) and $c_\lambda$. Here $\lambda_{\rm{crit}}\approx0.0031$ gives a plausible region of 0.994 credibility. (c)~Similarly, whenever Case~3 happens, the modifications result in $\lambda_{\rm{crit}}\approx0.0014$ for a plausible region of 0.99 credibility with a given dataset.}
\end{figure}

The value of $V_{\mathcal{M}^{(d=1)}_2}$ is simply equal to one, the Lebesgue length of the interval $0\leq r\leq1$. As the Lebesgue prior is defined for the entire $\mathcal{M}^{(d=1)}_2$, we have $L(\mathbb{D}|r=0)=L(\mathbb{D}|r=1)=0$ such that only Case~1 and 3 apply\footnote{In \cite{BR:part2}, where we study single-phase estimation with Bayesian regions for a different purpose, Case~2 shall apply to an enforced uniform prior that covers a subset of the phase interval since the likelihood at the boundary points can be nonzero in this case.}. Rejection sampling is certainly not necessary for such a simple class of states. Figure~\ref{fig:cred_1D} studies the behaviors of theoretical results for these two cases.

\begin{figure}[t]
	\center
	\includegraphics[width=1\columnwidth]{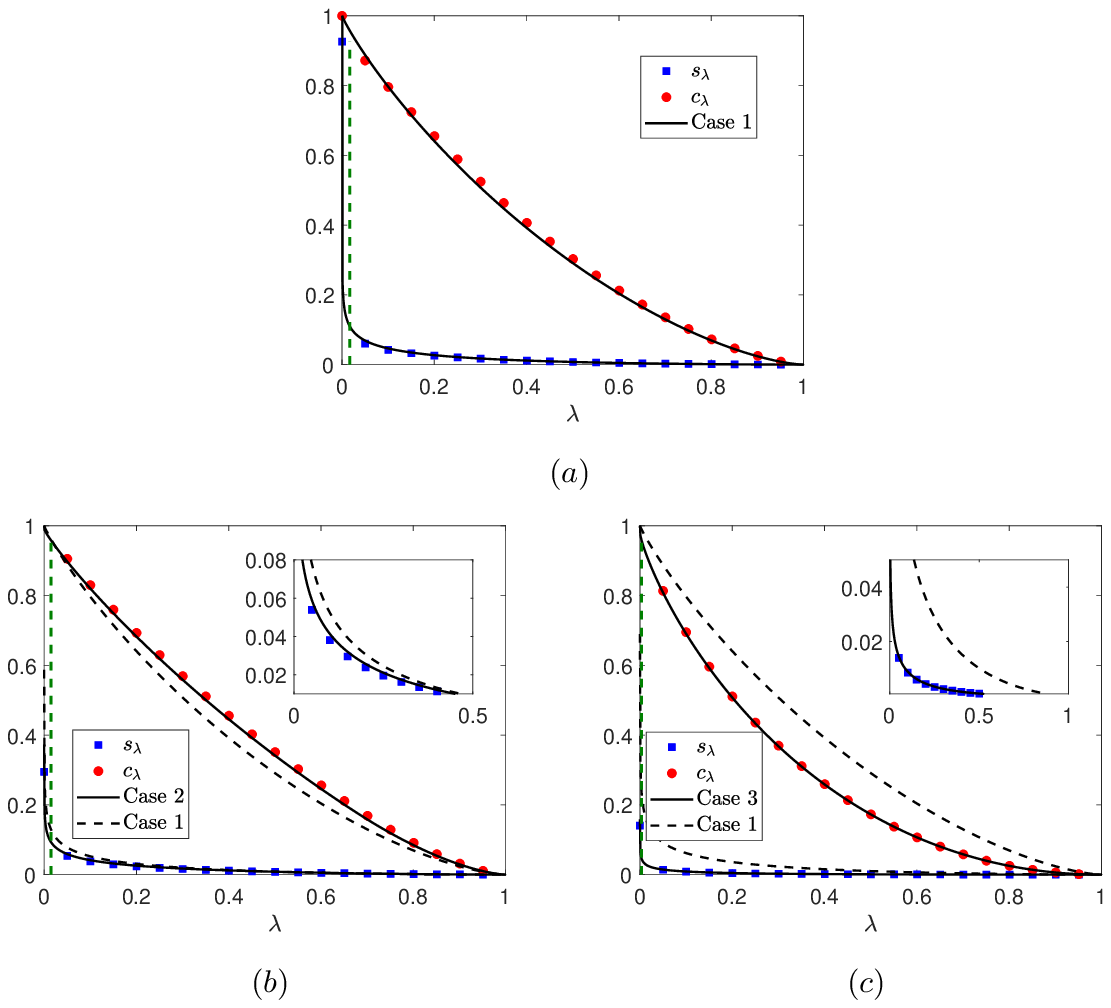}
	\caption{\label{fig:cred_3D}Full qubit estimation. Credible-region quantities are plotted for tomography on the complete qubit characterized by $\rvec{r}=(0.8,0.4,0.1)$ using the tetrahedron measurement by measuring data made up of $N=90$ copies. (a) In the optimistic Case~1, the plausible region, of 0.927 credibility, is defined by $\lambda_\mathrm{crit}\approx0.017$. (b)~With the same $N$, boundary effects begin to influence the characteristics of both region size and credibility when $\ML{\rvec{r}}$ is near $\partial\mathcal{R}\cap\partial\mathcal{M}_2$ as in Case~2, giving a plausible region of 0.963 credibility at $\lambda_\text{crit}\approx0.015$ for a particular dataset. (c)~Case~3 happens rather frequently as well, with an example dataset that gives a plausible region of 0.964 credibility at $\lambda_\text{crit}\approx0.0033$.}
\end{figure}

\subsubsection{Two-parameter qubit $(d=2)$}
If this time, we know that only $r_3=0$, then $\rho$ lies in the plane $(r_1-1/2)^2+r_2^2\leq1/4$. The volume $V_{\mathcal{M}^{(d=2)}_2}$ of this two-parameter subspace $\mathcal{M}^{(d=2)}_2$ can then be easily calculated to be
\begin{eqnarray}
V_{\mathcal{M}^{(d=2)}_2}=\int_{\left(r'_1-\frac{1}{2}\right)^2+r'^2_2\leq \frac{1}{4}}\D r'_1\,\D r'_2=\frac{\pi}{4}\,,
\end{eqnarray}
and the yield percentage through uniform rejection sampling for these states is therefore equal to $39.27\%$. The POM employed is the $M=4$ ``crosshair'' measurement consisting of projections onto the eigenstates of both Pauli operators $\sigma_z$ and $\sigma_x=\ket{+}\bra{+}-\ket{-}\bra{-}$:
\begin{eqnarray}
p_1=&\frac{1}{2}\opinner{0}{\rho}{0}=\frac{r_1}{2}\,,\quad\quad\,\,\,\, p_3=\frac{1}{2}\opinner{+}{\rho}{+}=\frac{1}{2}(1+2r_2) \\
p_2=&\frac{1}{2}\opinner{1}{\rho}{1}=\frac{1-r_1}{2}\,,\quad p_4=\frac{1}{2}\opinner{-}{\rho}{-}=\frac{1}{2}(1-2r_2)\,.
\end{eqnarray}
Figure~\ref{fig:cred_2D} illustrates the validity of our theory.

\subsubsection{Three-parameter qubit $(d=3)$}
For full qubit tomography, we require a minimum set of $M=2^2=4$-outcome informationally complete (IC) POM to completely characterize the qubit quantum state. One may consider the popular tetrahedron POM comprising the four symmetrically oriented measurement outcomes (symmetric IC POM or SIC~POM)
\begin{eqnarray}
\vec{a}_1\,\widehat{=}\,\frac{1}{\sqrt{3}}
\begin{pmatrix}
1\\
1\\
1
\end{pmatrix},\,\vec{a}_2\,\widehat{=}\,\frac{1}{\sqrt{3}}
\begin{pmatrix}
-1\\
-1\\
1
\end{pmatrix},\,\vec{a}_3\,\widehat{=}\,\frac{1}{\sqrt{3}}
\begin{pmatrix}
-1\\
1\\
-1
\end{pmatrix},\,\vec{a}_4\,\widehat{=}\,\frac{1}{\sqrt{3}}
\begin{pmatrix}
1\\
-1\\
-1
\end{pmatrix}\,.
\end{eqnarray}
This qubit POM as well as its extensions to higher dimensions constitute an optimal class of measurements in quantum information under certain conditions~\cite{Englert:2004aa,Durt:2008wt,Bent:2015aa}. The volume of the $\mathcal{M}_2$ under the Lebesgue prior can be shown to be $\pi/6$
either by setting $D=2$ in \eref{eq:leb_vol} or simply calculating the spherical volume
\begin{eqnarray}
V_{\mathcal{M}_2}=\int_{\left(r'_1-\frac{1}{2}\right)^2+r'^2_2+r'^2_3\leq \frac{1}{4}}\D r'_1\,\D r'_2\,\D r'_3=\frac{4}{3}\pi\left(\frac{1}{2}\right)^3=\frac{\pi}{6}\,.
\end{eqnarray}
The yield percentage for $\mathcal{M}_2$ is 13.09\%. The analyses of all three cases are described in Fig.~\ref{fig:cred_3D}.

\begin{figure}[t]
	\center
	\includegraphics[width=1\columnwidth]{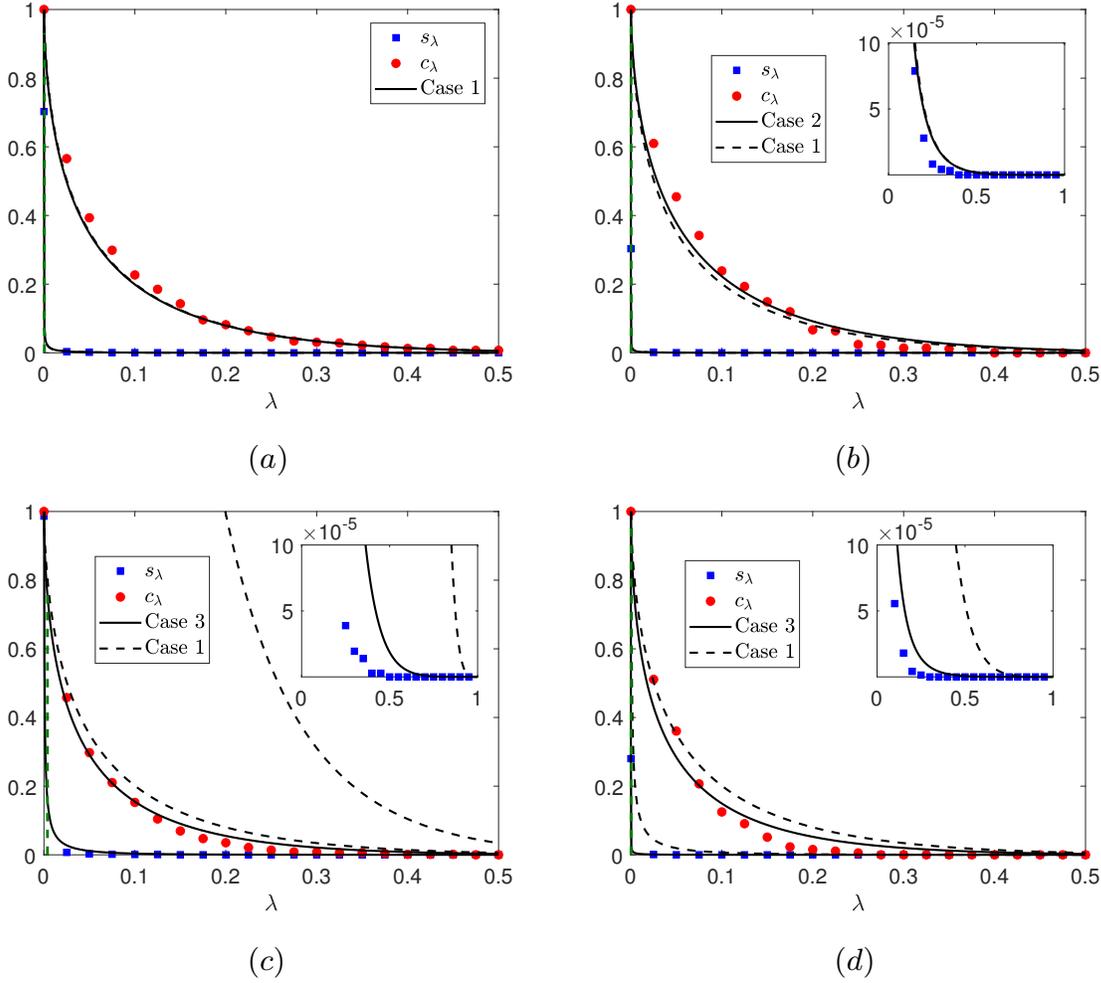}
	\caption{\label{fig:cred_8D} Qutrit Bayesian regions constructed with a $(M=90)$-outcome POM. (a)~Case~1 ($N=150$) and (b) Case~2 ($N=180$) are studied with the maximally-mixed true state $\rho=1/3$. (c,d) Case~3 refers to the true pure state described by the equal superposition $\ket{\,\,\,}=(\ket{0}+\ket{1}+\ket{2})/\sqrt{3}$ of three orthonormal kets. The 3rd case is presented with an ML estimator of (c)~rank-1 ($N=30$) and that for (d)~rank-2 ($N=90$). All insets blow up the scale for $s_\lambda$. Panels~(c) and (d) show that the (overesimated) size approximations still fare much better than the optimistic expressions in \eqref{eq:interior_cred}. Improvements on $s_\lambda$ estimates with asymptotic truncations become more conspicuous especially when (c) logarithmic divergence dominates in the low-$N$ regime, in which truncations can reduce a significant amount of Gaussian-approximation artifacts. Relevant values are found in the following table:\\
    \begin{tabular}{ccccc}
			\hline 
			Case  & Theory $\lambda_{\rm{crit}}$ & Theory $c_{\lambda_{\rm{crit}}}$ & Simulated $\lambda_{\rm{crit}}$ & Simulated $c_{\lambda_{\rm{crit}}}$\\
			1  & $5.52\times10^{-4}$ & 0.931 & $4.10\times10^{-4}$ & 0.972 \\
			2  & $1.55\times10^{-4}$ & 0.971 & $1.12\times10^{-4}$ & 0.988 \\
			3 (rank-1 ML)  & 0.0039 & 0.756 & $9.09\times10^{-4}$ & 0.938 \\
			3 (rank-2 ML) & $1.44\times10^{-4}$ & 0.953 & $6.58\times10^{-5}$ & 0.988 \\\hline
	\end{tabular}\\[10ex]}
\end{figure}

\subsection{Qutrit}

The qutrit is the next simplest quantum system of dimension $D=3$ which state 
\begin{eqnarray}
\rho\,\,\widehat{=}\begin{pmatrix}
r_1 & r_3+\I r_4 & r_5+\I r_6\\
r_3-\I r_4 & r_2 & r_7+\I r_8\\
r_5-\I r_6 & \,\,r_7-\I r_8\,\, & 1-r_1-r_2
\end{pmatrix}
\end{eqnarray}
can be completely characterized by the $(d=3^2-1=8)$-dimensional state parameter $\rvec{r}$. Therefore the minimum number of POM outcomes needed to estimate $\rvec{r}$ is $M=9$. The volume of the qutrit space, according to \eqref{eq:leb_vol}, is $V_{\mathcal{M}_3}=\pi^3/20160$. To compute $s_\lambda$ and $c_\lambda$ over $\mathcal{M}_3$, we may again perform uniform rejection sampling over the ranges $0\leq r_1,r_2\leq1$ and $-1\leq r_3,\ldots,r_8\leq1$. This time, we see that the yield percentage for $\mathcal{M}_3$ is significantly lower than that for $\mathcal{M}_2$---$2.4\times10^{-3}\,\%$ to be more precise for the uniform Lebesgue prior. Although it is possible to sample diagonal entries of $\rho$ such that $\tr{\rho}=1$ (\emph{i.e.} sampling on \emph{any} unit simplex) without sample wastage by renormalizing exponentially distributed random real numbers~\cite{Shang:2015mc,Seah:2015mc}, inevitably as $D$ grows, the method of rejection sampling for off-diagonal parameters rapidly becomes an inefficient and obsolete option for generating adequate parameter samples.

The qubit system possesses a dimension $D$ small enough such that the average error $\mathrm{E}[\norm{\rvec{r}-\ML{\rvec{r}}}]$ is small and the Gaussian approximations in \eqref{eq:gauss_approx} and \eqref{eq:gauss_approx2} are valid even when $N$ is not very large. Quantum systems of larger $D$, starting with the qutrit, generally requires a correspondingly larger $N$ to achieve similar tomographic precisions~\cite{Zhu:2014aa,Teo:2015qs}. For very large $N$ values, the likelihood function becomes extremely narrow since its curvature is asymptotically governed by $\FML\sim N$. As a result, the size $s_\lambda$ is tricky to calculate numerically with sophisticated Monte Carlo methods \cite{Shang:2015mc,Seah:2015mc}. For the purpose of demonstrating the performance of our results, we may slightly circumvent this problem by considering an overcomplete POM ($M>9$) while maintaining a reasonable $N$ value, which similarly reduces the average error~\cite{Zhu:2014aa} for the Gaussian approximations to hold. 

Figure~\ref{fig:cred_8D} showcases qutrit tomography for all the various cases discussed in Sec.~\ref{sec:results}. For qutrits, the size corrections are generally overestimates because of the complicated $\partial\mathcal{M}_3$.

\section{Conclusion}

We provided an asymptotic theory of Bayesian regions for general convex parameter spaces that cover a wide range of applications in quantum information whenever a uniform prior is used to describe the unknown true parameter. This allows any observer to conduct asymptotic error certification for uniform priors that avoids NP-hard Monte Carlo computations. The theory supplies analytical formulas for the region size and credibility in cases where the true parameter is an interior point [Eq.~\eqref{eq:interior_cred}, \eqref{eq:interior_plaus}, \eqref{eq:case2_1} and \eqref{eq:case2_2}], as well as the case where the true parameter is on the boundary of the parameter space [Eq.~\eqref{eq:case3_1} and \eqref{eq:case3_2}]. These expressions approach the exact answers whenever the joint boundary of both the region and full parameter space is smooth. Otherwise they generally give conservative overestimates for the region size as this is related to the way region truncations are handled by the theory. When applied to examples in quantum-state tomography, these asymptotic expressions give extremely accurate estimates in spite of the sophisticated state-space boundaries. The theoretical framework presented here can in principle be generalized to any other prior so long as analytical integrals for Gaussian likelihoods and the volume of the parameter space are known for that prior. This, however, has to be done on a case-by-case basis at the moment. 

\section{Acknowledgments}
We acknowledge financial support from the BK21 Plus Program (21A20131111123) funded by the Ministry of Education (MOE, Korea) and National Research Foundation of Korea (NRF), the NRF grant funded by the Korea government (MSIP) (Grant No. 2010-0018295), and the Korea Institute of Science and Technology Institutional Program (Project No. 2E27800-18-P043).

\appendix

\section{The derivation of \eqref{eq:interior_cred}}
\label{app:cs_int1}

We start with \eqref{eq:sc_def} and the Gaussian approximation in \eqref{eq:gauss_approx} for an interior ML estimator to first calculate the credible-region size. We proceed by using the well-known integral representation
\begin{eqnarray}
\eta(x)=\left.\int\dfrac{\D t}{2\pi\I}\,\dfrac{\E{\I xt}}{t-\I\epsilon}\right|_{\epsilon=0}
\end{eqnarray}
of the Heaviside step function and the recognition that $\eta(L(\mathbb{D}|\rvec{r})-\lambda L_\text{max})=\eta(\log L(\mathbb{D}|\rvec{r})-\log\left(\lambda L_\text{max}\right))$ to write
\begin{eqnarray}
s_\lambda&=&\,\int_{\mathcal{R}_0}(\D\,\rvec{r}')\,\chi_\lambda(\rvec{r}')\,,\nonumber\\
&=&\,\left.\int\dfrac{\D t}{2\pi\I}\,\dfrac{\E{-\I t\log\left(\lambda L_\text{max}\right)}}{t-\I\epsilon}\,\int_{\mathcal{R}_0}(\D\,\rvec{r}')\,\E{\I t\log L(\mathbb{D}|\rvec{r}')}\right|_{\epsilon=0}\,,
\end{eqnarray}
where after a reminder that $(\D\,\rvec{r}')$ is a normalized measure, the integral in $\rvec{r}'$ can be simplified to
\begin{eqnarray}
\int_{\mathcal{R}_0}(\D\,\rvec{r}')\,\E{\I t \log L(\mathbb{D}|\rvec{r}')}&\approx&\,\dfrac{\E{\I t\log L_\text{max}}}{V_{\mathcal{R}_0}}\int\left(\prod_j\D\,r'_j\right)\,\E{-\frac{\I t}{2}(\rvec{r}'-\ML{\rvec{r}})\bm{\cdot}\FML\bm{\cdot}(\rvec{r}'-\ML{\rvec{r}})}\nonumber\\
&=&\,\dfrac{\E{\I t\log L_\text{max}}}{V_{\mathcal{R}_0}}\left(\dfrac{2\pi}{\I t}\right)^{d/2}\left(\DET{\FML}\right)^{-1/2}\,.
\end{eqnarray}
The integral in $t$ can then be completed with another identity
\begin{eqnarray}
\dfrac{1}{a^n}=\dfrac{1}{(n-1)!}\int^\infty_0\D y\,y^{n-1}\E{-ay}\,:
\end{eqnarray}
\begin{eqnarray}
s_\lambda&=&\,\dfrac{(2\pi)^{d/2}}{V_{\mathcal{R}_0}}\left(\DET{\FML}\right)^{-1/2}\left.\int\dfrac{\D t}{2\pi\I}\,\dfrac{\E{-\I t\log\lambda }}{(\I t)^{d/2}\,(t-\I\epsilon)}\right|_{\epsilon=0}\nonumber\\
&=&\,\dfrac{(2\pi)^{d/2}}{V_{\mathcal{R}_0}\,(d/2-1)!}\left(\DET{\FML}\right)^{-1/2}\underbrace{\int^\infty_0\D y\,y^{d/2-1}\left.\int\dfrac{\D t}{2\pi\I}\,\dfrac{\E{-\I t(\log\lambda+y) }}{t-\I\epsilon}\right|_{\epsilon=0}}_{\qquad\qquad\qquad\quad\displaystyle{=\int^{-\log\lambda}_0\D y\,y^{d/2-1}}}\nonumber\\
&=&\,\dfrac{V_d}{V_{\mathcal{R}_0}}(-2\log\lambda)^{d/2}\left(\DET{\FML}\right)^{-1/2}\,.
\end{eqnarray}

The credibility may be calculated either with \eqref{eq:sc_def} or \eqref{eq:cs_relation}. We choose the latter route as an example, along which we need the ingredients
\begin{eqnarray}
\int^1_\lambda\D\lambda'\,(-\log\lambda')^{\alpha}&=&\,\alpha\,!-\Gamma\left(\alpha+1,-\log\lambda\right)\,,\nonumber\\
\qquad\quad\Gamma(\alpha+1,y)&=&\,\alpha\,\Gamma(\alpha,y)+y^\alpha\E{-y}
\end{eqnarray}
for the upper incomplete Gamma function. A little algebraic manipulation after that leads to the answer.

\section{The estimation of $\rvec{r}_P$}
\label{app:case2_rP}

As $\ML{\rvec{r}}$ is close to $\partial\mathcal{R}\cap\partial\mathcal{R}_0$, the column $\rvec{r}_P$ can be estimated by first generating a set $\left\{\rvec{r}^{(\text{bd})}_j\right\}^L_{j=1}$ of $L$ boundary parameter columns, which can be done by generating many random $d$-dimensional columns $\rvec{\epsilon}_j$ of small magnitudes and defining $\rvec{r}^{(\text{bd})}_j=\mathcal{M}(\ML{\rvec{r}}+\rvec{\epsilon}_j)$, where $\mathcal{M}$ is a map that brings any column that lies outside of $\mathcal{R}_0$ to $\partial\mathcal{R}_0$ (the probability of generating a random boundary point without the action of $\mathcal{M}$ is effectively zero). Then $\rvec{r}_P$ may be taken to be the boundary point that gives the maximal likelihood value $L^{(\partial\mathcal{R}_0)}_\text{max}$.

As an example, we suppose that in state tomography, $\ML{\rvec{r}}$ is the $(d=D^2-1)$-dimensional real parameter column that uniquely represents the $D$-dimensional ML quantum state $\ML{\rho}$ that lies close to $\partial\mathcal{R}\cap\partial\mathcal{R}_0$. Then a set of random columns $\rvec{\epsilon}_j$, distributed according to the standard Gaussian distribution for instance, is added to $\ML{\rvec{r}}$ one at a time and the resulting columns $\ML{\rvec{r}}+\rvec{\epsilon}_j\rightarrow H_j$ are transformed into the corresponding Hermitian operators $H_j=H_j^\dagger$. We discard those $H_j$s that are full-rank positive operators and move on to others that are nonpositive, and apply the map $\mathcal{M}(\cdot)=\mathcal{N}[\,\,\cdot+\sigma_\text{min}(\cdot)1]$ to $H_j$, which adds a multiple of the identity equal to the minimum eigenvalue $\sigma_\text{min}$ and trace-normalize the resulting operator. This turns the nonpositive $H_j$s into boundary states $\rho^{(\text{bd})}_j\rightarrow\rvec{r}^{(\text{bd})}_j$ that is near $\ML{\rvec{r}}$ if $\rvec{\epsilon}_j$ is small enough.

\section{The derivation of \eqref{eq:case2_1}}
\label{app:case2_deriv}

With the Gaussian likelihood in \eqref{eq:gauss_approx} centered at $\ML{\rvec{r}}$, let us denote the full hyperellipsoid defined by the isolikelihood contour at some value of $\lambda$ as $\mathcal{E}_\lambda$. If $\mathcal{R}=\mathcal{R}_\lambda$ is truncated, then the region $\widetilde{\mathcal{R}}_\lambda\supseteq\mathcal{R}_{\lambda}$ that is bounded $\partial\mathcal{E}_{\lambda}\cap\partial P$ is an overestimate of $\mathcal{R}_{\lambda}$. The task here is to calculate the volume $V_{\widetilde{\mathcal{R}}_{\lambda}}$ of this region.

The hyperellipsoidal surface $\partial\mathcal{E}_\lambda$ for any $\lambda$ is described by the equation
\begin{eqnarray}
(\rvec{r}-\ML{\rvec{r}})\bm{\cdot}\FML'\bm{\cdot}(\rvec{r}-\ML{\rvec{r}})=1
\end{eqnarray}
with $\FML'=\FML/\left(-2\log\lambda\right)$, or in terms of its more convenient diagonal-basis representation found with the spectral decomposition $\FML'=\dyadic{O}\dyadic{D}\,\TP{\dyadic{O}}$,
\begin{eqnarray}
(\rvec{r}'-\ML{\rvec{r}}')\bm{\cdot}\dyadic{D}\bm{\cdot}(\rvec{r}'-\ML{\rvec{r}}')=1\,,
\end{eqnarray}
where $\rvec{a}'=\TP{\dyadic{O}}\bm{\cdot}\rvec{a}$, where the diagonal entries $D_j$ of $\dyadic{D}$ are reciprocals of squares of the $\lambda$-hyperellipsoidal axes lengths. In the primed coordinates, the hyperplane $P$, which contains $\rvec{r}'_P$, the ML estimator over $\partial\mathcal{R}_0$, and the normal $\rvec{n}'\propto\dyadic{D}\bm{\cdot}\left(\rvec{r}'_P-\ML{\rvec{r}}'\right)$, satisfies the equation $\rvec{n}'\bm{\cdot}\rvec{r}'=\rvec{n}'\bm{\cdot}\rvec{r}'_P$. One easy trick to calculate $V_{\widetilde{\mathcal{R}}_{\lambda}}$ would then be to first start with the integral definition
\begin{eqnarray}
V_{\widetilde{\mathcal{R}}_{\lambda}}=V_{\mathcal{R}_0}\int(\D\,\rvec{r}')\,\eta\left(1-(\rvec{r}'-\ML{\rvec{r}}')\bm{\cdot}\dyadic{D}\bm{\cdot}(\rvec{r}'-\ML{\rvec{r}}')\right)\eta\left(\rvec{n}'\bm{\cdot}(\rvec{r}'_P-\rvec{r}')\right)\,,
\label{eq:Vintersect}
\end{eqnarray}
and next perform the change of variables $\rvec{r}'\rightarrow \rvec{r}''=\dyadic{D}^{1/2}\bm{\cdot}(\rvec{r}'-\ML{\rvec{r}}')$ to express this same volume
\begin{eqnarray}
V_{\widetilde{\mathcal{R}}_{\lambda}}=\dfrac{V_{\mathcal{R}_0}}{\sqrt{\DET{\dyadic{D}}}}\int_{S_{d-1}}(\D\,\rvec{r}'')\,\eta\left(\rvec{n}'\bm{\cdot}\left(\rvec{r}'_P-\ML{\rvec{r}}'\right)-\rvec{n}'\bm{\cdot}\dyadic{D}^{-1/2}\bm{\cdot}\rvec{r}''\right)
\end{eqnarray}
as a multiple of the volume of intersection between a corresponding unit \mbox{$(d-1)$}-hypersphere $S_{d-1}$ and a transformed hyperplane $P'$ described by the equation \mbox{$\rvec{n}'\bm{\cdot}\dyadic{D}^{-1/2}\bm{\cdot}\rvec{r}''=\rvec{n}'\bm{\cdot}\left(\rvec{r}'_P-\ML{\rvec{r}}'\right)$} in the $\rvec{r}''$ reference frame. 

For the primitive prior and the earlier definition of $\rvec{n}'$, this intersection volume has a known analytical answer, which depends on the shortest distance
\begin{eqnarray}
l&=&\,l_0=\dfrac{\left|\rvec{n}'\bm{\cdot}\left(\rvec{r}'_P-\ML{\rvec{r}}'\right)\right|}{\norm{\dyadic{D}^{-1/2}\bm{\cdot}\rvec{n}'}}\nonumber\\
&=&\,\sqrt{\dfrac{\left(\rvec{r}_P-\ML{\rvec{r}}\right)\bm{\cdot}\FML\bm{\cdot}\left(\rvec{r}_P-\ML{\rvec{r}}\right)}{-2\log\lambda}}=\sqrt{\dfrac{\log \lambda_\text{int}}{\log\lambda}}
\end{eqnarray}
between the center of the hypersphere and $P'$. It follows that the magnitude of $l_0$ increases with $\lambda$. At the critical value $\lambda=\lambda_\text{int}$, we have $l_0=1$, which tells us that at this critical value $\partial\mathcal{E}_{\lambda\geq\lambda_\text{int}}\cap\partial P=\emptyset$. Beyond $\lambda>\lambda_\text{int}$ we must have the shortest distance $l=1$ set to unity since this would imply that $V_{\widetilde{\mathcal{R}}_0}=V_{\mathcal{E}_{\lambda}}=\gamma V_{d,{\lambda}}$. It can then be shown, for instance see \cite{Li:2011av}, that $V_{\widetilde{\mathcal{R}}_{\lambda}}=\gamma V_{\mathcal{E}_{\lambda}}=\gamma V_{d,{\lambda}}$, where $\gamma=1-\BETAr{\frac{1-l}{2}}{\dfrac{d+1}{2},\dfrac{d+1}{2}}$.

\section*{References}
\bibliography{Bayes.reg.1}

\end{document}